\documentclass[reprint,prd,aps,a4paper]{revtex4-1}

\usepackage{amssymb}
\usepackage{amsmath}
\usepackage{bbold}
\usepackage{graphicx}
\usepackage{hyperref}
\usepackage{afterpage}

\newcommand{\p}{p_{new}}
\newcommand{\ere}{\langle D \rangle}

\begin{document}

\author{Sebastian Seehars}
\email{seehars@phys.ethz.ch}
\author{Adam Amara}
\author{Alexandre Refregier}
\author{Aseem Paranjape}
\author{Jo\"el Akeret}
\affiliation{ETH Zurich, Department of Physics, Wolfgang-Pauli-Strasse 27, 8093 Zurich, Switzerland}

\title{Information Gains from Cosmic Microwave Background Experiments}

\begin{abstract}
To shed light on the fundamental problems posed by Dark Energy and Dark Matter, a large number of experiments have been performed and combined to constrain cosmological models. We propose a novel way of quantifying the information gained by updates on the parameter constraints from a series of experiments which can either complement earlier measurements or replace them. For this purpose, we use the Kullback-Leibler divergence or relative entropy from information theory to measure differences in the posterior distributions in model parameter space from a pair of experiments. We apply this formalism to a historical series of Cosmic Microwave Background experiments ranging from Boomerang to WMAP, SPT, and Planck. Considering different combinations of these experiments, we thus estimate the information gain in units of bits and distinguish contributions from the reduction of statistical errors and the `surprise' corresponding to a significant shift of the parameters' central values. For this experiment series, we find individual relative entropy gains ranging from about 1 to 30 bits. In some cases, e.g. when comparing WMAP and Planck results, we find that the gains are dominated by the surprise rather than by improvements in statistical precision. We discuss how this technique provides a useful tool for both quantifying the constraining power of data from cosmological probes and detecting the tensions between experiments.
\end{abstract}

\maketitle

\section{Introduction} 
\label{sec:introduction}

\noindent Over recent decades, observational evidence in support of the $\Lambda$CDM model has grown steadily. Though some of the key ingredients of the model, including Dark Matter and Dark Energy, are not fully understood, an impressive array of new experiments show findings consistent with predictions of the model. Chief among the datasets are high-precision measurements of the Cosmic Microwave Background (CMB) \cite{Wright:1996um,MacTavish:2008tm,Bennett:2012wy,Hinshaw:2012vg,Story:2012vh,Das:2011gd}. This area has received significant attention recently with the release of the first cosmological analysis of data from the Planck satellite \cite{Collaboration:2013ww,Collaboration:2013uv}. This experiment can be seen as the latest in a long line of measurements that have targeted the CMB. At each step, data has been used to place constraints on the parameters of the $\Lambda$CDM model using Bayesian inference to represent the constraints as a probability density in parameter space called the posterior distribution. To judge the progress made between successive measurements, a framework for comparing probability distributions is needed.

One method to quantify the difference between the constraints from different surveys is the relative entropy or Kullback-Leibler divergence \cite{Kullback:1951va} between the respective distributions. Initially motivated from information theory, relative entropy has been proposed in the cosmology literature for forecasting and experiment design \cite{Paykari:2013dd,Amara:2013wv,March:2011ij,0004-637X-771-1-12} as well as for parameter estimation and model selection \cite{Verde:2013hp,Kunz:2006de}. In this paper, the relative entropy is introduced as a new tool for measuring the information gained from individual experiments by applying it to their posteriors on the full cosmological parameter space. Two distinct cases of data combinations are analyzed: adding complementary data to existing constraints and replacing data with a more accurate but correlated measurement.

The relative entropy between two posteriors measures gains in statistical precision and shifts of confidence regions at the same time. Disentangling these contributions is of great interest for detecting tensions between datasets. In the limit of linear models and Gaussian likelihoods, it is shown that the relative entropy can indeed be separated into an expected part measuring the improvements in precision and a contribution from shifts in the distribution means that is named `surprise'. Explicit expressions for the relative entropy and its decomposition into expected relative entropy and surprise are derived in this limit and can be evaluated from moments of the posteriors.

These concepts are then applied to the posteriors of the $\Lambda$CDM parameters from the Balloon Observations of Millimetric Extragalactic Radiation and Geophysics (BOOMERANG) \cite{MacTavish:2008tm}, the Wilkinson Microwave Anisotropy Probe (WMAP) \cite{Bennett:2012wy,Hinshaw:2012vg}, the South Pole Telescope (SPT) \cite{Story:2012vh}, and the Planck \cite{Collaboration:2013ww,Collaboration:2013uv} CMB surveys. Using the Monte Carlo Markov chain framework \verb|CosmoHammer| \cite{Akeret:2012uk}, estimates for the relative entropy, its expected, and its surprise contributions are given for different combinations of these datasets. The concepts can be easily applied to other probes, too.

This paper is organized as follows. In section \ref{sec:information_in_parameter_estimation} the connection between relative entropy and parameter estimation is discussed. The results for disentangling expected relative entropy and surprise in the Gaussian limit are derived in section \ref{sec:the_gaussian_limit}. Applying those concepts to CMB surveys, numerical results for the relative entropy between BOOMERANG, WMAP, SPT, and Planck data are shown in section \ref{sec:information_gains_from_cmb}. The conclusions are summarized in section \ref{sec:discussion}.


\section{Relative Entropy in Cosmological Parameter Estimation} 
\label{sec:information_in_parameter_estimation}

\noindent In this section, the application of relative entropy to cosmological parameter estimation as a measure for information gains from astronomical observations is discussed after a short introduction to both concepts.

\subsection{Cosmological parameter estimation} 
\label{sub:parameter_estimation}

\noindent A typical problem in cosmology is the inference of the parameters of a cosmological model from astronomical data. In most cases this amounts to comparing observables, such as correlation functions or power spectra, that can be both measured by surveys and predicted from a model. Given such observables and a model with parameters $\Theta = (\theta_1, \cdots, \theta_d)$, one can typically construct a likelihood function for the parameters, i.e. the probability distribution of the data $\mathcal D$ given the model parameters $\Theta$:
\begin{equation}
	\mathcal L(\Theta;\mathcal D) \equiv p(\mathcal D|\Theta).
\end{equation}
When prior information on the model parameters is available in the form of a probability density $p(\Theta)$, Bayes' theorem describes how to update the knowledge on $\Theta$ by accounting for the data:
\begin{equation}
		\p(\Theta) \equiv p(\Theta|\mathcal D)
		= \frac {\mathcal L(\Theta;\mathcal D) p(\Theta)} {\int d\Theta\, \mathcal L(\Theta;\mathcal D) p(\Theta)},
		\label{eq:bayest}
\end{equation}
where $\p$ is called the posterior distribution of the parameters. The denominator is often called the evidence, $E(\mathcal D)$, of the data and is equivalent to the distribution of the data anticipated from prior knowledge on the model, evaluated at the actual measurement:
\begin{equation}
	\begin{aligned}
		E(\mathcal D) &\equiv \int d\Theta\, \mathcal L(\Theta;\mathcal D)p(\Theta)
		\\&= \int d\Theta\, p(\mathcal D|\Theta)p(\Theta) = p(\mathcal D).
	\end{aligned}
	\label{eq:evidence}
\end{equation}
The application of relative entropy to the problem of quantifying changes in the posterior is discussed next.

\subsection{Relative entropy} 
\label{sub:relative_entropy}

\noindent First defined by Kullback and Leibler in 1951 \cite{Kullback:1951va}, relative entropy or Kullback-Leibler divergence is an important concept in information theory. It aims at measuring differences between two probability densities. In data compression, for example, it has a clear interpretation as the number of extra bits needed when assuming a wrong distribution of the data's alphabet (for more details, see \cite{Cover} for example).

In cosmology, the relative entropy has been proposed as a tool for experiment design, forecasting, and model selection. \citet{March:2011ij} constructed a figure of merit that is related to the relative entropy in order to study the robustness of parameter constraints to possible systematic errors. \citet{Paykari:2013dd} employed a special case of the relative entropy to forecast the constraints from different survey strategies with and without sparse sampling of the sky. \citet{Amara:2013wv} used the relative entropy between distributions in data space to compare the model breaking potential of different surveys. \citet{Kunz:2006de} applied a Bayesian measure of complexity related to relative entropy to the problem of model selection in cosmology. \citet{Verde:2013hp} used relative entropy to study the one dimensional marginals of WMAP and Planck constraints for the parameters of $\Lambda$CDM and extensions to the basic model.

This paper focuses on applying the relative entropy to the full multivariate posterior distributions in order to develop a new tool for comparing the constraints from different datasets. In order to define relative entropy, let $X$ be a continuous, $d$ dimensional random variable with probability density functions $p_1(X)$ and $p_2(X)$. The relative entropy $D(p_2||p_1)$ between $p_2$ and $p_1$ is then given by
\begin{equation}
	D(p_2||p_1) \equiv \int_\mathcal S dX\, p_2(X) \log \frac {p_2(X)} {p_1(X)},
	\label{eq:relativeentropy}
\end{equation}
where $\mathcal S$ is the support of $p_2$. Note that the base of the logarithm in equation \eqref{eq:relativeentropy} has to be chosen to equal $2$ in order to interpret $D(p_2||p_1)$ as an information gain measured in bits. When evaluated for the natural logarithm the unit is called nats and can be simply transformed into bits by dividing $D$ by $\log(2)$. $D$ is finite only if $\mathcal S$ is contained in the support of $p_1$. Although not being symmetric in $p_1$ and $p_2$, $D(p_2||p_1)$ is often interpreted as a distance between the two distributions as it is non-negative, $D(p_2||p_1) \geq 0 $, and zero if and only if $p_1 = p_2$ almost everywhere \cite{Cover}. It is furthermore easy to see that the relative entropy is invariant under invertible transformations in $X$: as probability distributions satisfy $p(Y)=p(X)\left|dX/dY\right|$ for an invertible transformation $Y(X)$, the Jacobian matrices $dX/dY$ cancel in the logarithm and $p_2(X)dX=p_2(Y)dY$. Finally, the relative entropy is additive if $X$ can be decomposed into independent sets of variables for both $p_1$ and $p_2$ \cite{Kullback:1951va}. The following section introduces relative entropy as a diagnostic in two important cases of updating the constraints on cosmological parameters.

\subsubsection{Adding complementary data} 
\label{ssub:adding_supplementary_data}

\noindent Consider the case of sequential updating of the parameter constraints with uncorrelated or very weakly correlated datasets that complement each other. As an example, one might think of updating the constraints from a CMB survey with supernova data or from low-$\ell$ multipoles of the CMB power spectrum as measured by WMAP with high-$\ell$ data from SPT. The problem is hence to update prior knowledge $p(\Theta)$ from one dataset using the new data with likelihood $\mathcal L(\Theta)$ via equation \ref{eq:bayest}. When focusing on measuring information gains in such a sequential updating scheme, the quantity of interest is given by the relative entropy between prior $p$ and posterior $\p$, defined by
\begin{equation}
	D(\p||p) = \int d\Theta\, \p(\Theta) \log \frac {\p(\Theta)} {p(\Theta)}.
	\label{eq:relentpostprior}
\end{equation}
$D(\p||p)$ quantifies the difference between the parameter distributions before and after updating with the new data. It can hence be interpreted as a measure for the amount by which the constraints on the model have to be changed when accounting for the new data. Due to the invariance under parameter transformations of the relative entropy, this measure does not depend on the particular parametrization of a given model.


\subsubsection{Replacing data} 
\label{ssub:substituting_data}

\noindent Another important case is parameter estimation from correlated datasets. A typical example of correlated datasets in cosmology are power spectra on large scales because of cosmic variance. Obviously, the likelihood of two correlated datasets cannot be described by two independent functions, preventing a joint analysis with Bayes' theorem as in section \ref{ssub:adding_supplementary_data}.

Whenever a new dataset is strongly correlated but superior to old data, thinking of BOOMERANG and WMAP CMB power spectra for example, a joint analysis is typically too complex compared to the expected effects on the precision. The more accurate new data is then usually simply used to replace the correlated older data in the parameter estimation step. But there are also more complex situations of parameter estimation from correlated datasets, for example a partial replacement of data or a joint likelihood function for both datasets that correctly takes correlations into account. An example for the former is the replacement of the WMAP temperature power spectrum with new Planck data, while the latter case can be illustrated with the successive WMAP releases after one, three, five, seven, and nine years of data collection. In any of these cases, the relative entropy between the two separately analyzed datasets is a useful diagnostic for measuring changes in the posteriors and detecting inconsistencies.

Usually starting from the same prior $p$, the posteriors from the old likelihood $\mathcal L_{old}(\Theta|D_{old})$ and the new likelihood $\mathcal L_{new}(\Theta|D_{new})$ result in the posteriors $p_{old}$ and $\p$ as given by equation \eqref{eq:bayest}. The quantity of interest in this case is given by
\begin{equation}
	D(\p||p_{old}) = \int d\Theta\, \p(\Theta) \log \frac {\p(\Theta)} {p_{old}(\Theta)}.
	\label{eq:relentpostprior2}
\end{equation}
$D(\p||p_{old})$ quantifies the difference between the constraints coming from the two datasets alone and is therefore a measure for shifts in the confidence regions as well as for improvements in precision. In general, disentangling those contributions is hard, but for the limit where all distributions are Gaussian, some useful results are shown next.



\section{Gaussian Limit} 
\label{sec:the_gaussian_limit}

\noindent In this section, an analytically tractable example for parameter estimation is considered that turns out to be useful when analyzing CMB data. The likelihood is modeled as a normal distribution in the data $\mathcal D$ centered around the model predictions $F(\Theta)$ with a fixed data covariance $\mathcal C$:
\begin{equation}
	\mathcal L(\Theta;\mathcal D) = \mathcal N(\mathcal D;F(\Theta),\mathcal C),
	\label{eq:normlike}
\end{equation}
where $\mathcal N(x;\mu,\Sigma)$ hereafter denotes a multivariate normal distribution in $x$ with mean $\mu$ and covariance $\Sigma$, the parameters $\Theta$ are of dimensionality $d$, and the data vector $\mathcal D$ has $n > d$ dimensions. Furthermore, a normally distributed prior, i.e. $p(\Theta)=\mathcal N(\Theta;\Theta_p,\Sigma_p)$, and a model which is linear in $\Theta$ is considered:
\begin{equation}
	F(\Theta) = F_0 + M\Theta.
	\label{eq:linmod}
\end{equation}
Under these assumptions, the posterior $\p$ is also normally distributed in $\Theta$ (see Appendix \ref{sub:deriving_the_posterior} for more details):
\begin{align}
	\p(\Theta) &= \mathcal N (\Theta; \Theta_{new}, \Sigma_{new}),\label{eq:ggpost}\\
	\Sigma_{new} &= \left(\Sigma_p^{-1} + M^T\mathcal C^{-1}M\right)^{-1},\\
	\Theta_{new} &= \Sigma_{new} \left(\Sigma_p^{-1}\Theta_p + M^T\mathcal C^{-1}(\mathcal D - F_0) \right).
\end{align}
Note that the Fisher matrix of the likelihood \cite{Fisher:1925gt} in this limit is given by
\begin{equation}
	\begin{aligned}
		I_{ij}(\Theta) &\equiv \int d\mathcal D\, \mathcal L(\Theta;\mathcal D) \left( \frac {\partial \log \mathcal L} {\partial \theta_i} \right)\left( \frac {\partial \log \mathcal L} {\partial \theta_j} \right)\\&=M^T\mathcal C^{-1}M,
	\end{aligned}
	\label{eq:fi}
\end{equation}
which also appears in $\Sigma_{new}$. Although linearity of the model and Gaussianity of both likelihood and prior are strong requirements, these conditions turn out to be reasonable approximations for most of the CMB data analysis as is demonstrated in section \ref{sec:information_gains_from_cmb}. The relative entropy between two Gaussians $p_1(\Theta) = \mathcal N(\Theta;\Theta_1,\Sigma_1)$ and $p_2(\Theta) = \mathcal N(\Theta;\Theta_2,\Sigma_2)$ is well known (see appendix \ref{sub:relative_entropy_between_gaussians}) and is given by
\begin{equation}
	\begin{aligned}
		&D(p_2||p_1) = \frac 1 2 (\Theta_1 - \Theta_2)^T \Sigma_1^{-1} (\Theta_1 - \Theta_2)\\
		&\qquad + \frac 1 2 \left(\text{tr}(\Sigma_2\Sigma_1^{-1}) -d - \log \frac {\det(\Sigma_2)} {\det(\Sigma_1)} \right).
	\end{aligned}
	\label{eq:ggre}
\end{equation}
The same relation is also used in \citet{Amara:2013wv}, while \citet{Paykari:2013dd} and \citet{March:2011ij} restrict themselves to aligned means. As can be seen from equation \eqref{eq:ggre}, the relative entropy contains ratios between the covariance matrices $\Sigma_2$ and $\Sigma_1$ as well as a weighted difference between their means $\Theta_1$ and $\Theta_2$. From a parameter estimation point of view, this can be intuitively understood as contributions from an increase in the precision of the measurement and from the significance of the shifts in the central values of the constraints, respectively. Next, this distinction is made more explicit by separating the relative entropy into an expected and a surprise part with the former measuring gains in precision and the latter quantifying the significance of the shifts in parameter space. As in section \ref{sub:relative_entropy}, the cases of adding complementary data and replacing correlated data are considered in the following.

\subsection{Adding complementary data} 
\label{sub:adding_supplementary_data2}

\noindent As introduced in section \ref{ssub:adding_supplementary_data}, the quantity of interest when updating the constraints of one survey with complementary results from another is the difference between prior and posterior knowledge. In the Gaussian limits discussed in this section, $D(\p||p)$ amounts to
\begin{equation}
	\begin{aligned}
		&D(\p||p) = \frac 1 2 (\Theta_{new} - \Theta_p)^T\Sigma_p^{-1}(\Theta_{new} - \Theta_p)\\
		&\qquad +\frac 1 2 \left(\text{tr}\left(\Sigma_p^{-1}\Sigma_{new}\right) - d - \log \frac {\det \Sigma_{new}} {\det\Sigma_p}\right).
	\end{aligned}
	\label{eq:regg}
\end{equation}
$D(\p||p)$ can be seen as a function of the data used to derive $\p$, hereafter denoted as $\mathcal D_{new}$. At the same time, the observed $\mathcal D_{new}$ is expected to be a realization from the evidence $E$ as defined in equation \eqref{eq:evidence}, as $E$ is the prior distribution for the data $\mathcal D_{new}$. Hence, the prior distribution for $\mathcal D_{new}$ induces a distribution for $D(\p||p)$ which is discussed in Appendix \ref{sub:expected_relative_entropy}. The expected relative entropy $\ere$ is then defined as the mean value of the prior distribution of $D(\p||p)$:
\begin{equation}
	\ere \equiv \int d\mathcal D_{new}\, D(\p||p)E(\mathcal D_{new}).
\end{equation}
In Appendix \ref{sub:expected_relative_entropy} it is shown that under the assumptions of this section, $\ere$ is given by
\begin{equation}
	\ere = -\frac 1 2 \log \frac {\det \Sigma_{new}}{\det\Sigma_p}.
	\label{eq:expre}
\end{equation}
In a similar fashion, the standard deviation of the expected relative entropy $\sigma(D)$ can be evaluated:
\begin{equation}
	\begin{aligned}
		\sigma^2(D) &\equiv \int d\mathcal D_{new}\, (D-\ere)^2 E(\mathcal D_{new})\\
		&= \frac 1 2 \text{tr}\left((\Sigma_p^{-1}\Sigma_{new} - \mathbb 1)^2\right).
	\end{aligned}
	\label{eq:sigmad}
\end{equation}
Again, the second line is true for the special case considered in this section. The surprise $S$ is then defined as the difference between observed and expected relative entropy:
\begin{equation}
	S \equiv D(\p||p) - \ere.
	\label{eq:surprise}
\end{equation}
As only the mean $\Theta_{new}$ of $\p$ depends on $\mathcal D_{new}$, the surprise actually quantifies the difference between the expected mean shift $\langle(\Theta_{new} - \Theta_p)^T\Sigma_p^{-1}(\Theta_{new} - \Theta_p)\rangle$ and the observed shift. The surprise $S$ is anticipated to be of order $\sigma(D)$. However, the distribution of the relative entropy $D$ is actually a generalized chi-squared distribution and $\sigma(D)$ does not give full information on the significance of deviations from $\ere$. The p-values for $S$, i.e. the prior probability of a surprise that is greater or equal than the observed surprise $S$, can be calculated numerically as shown in Appendix \ref{sub:expected_relative_entropy}.


\subsection{Replacing data} 
\label{sub:substituting_data2}

\noindent Following the arguments from section \ref{ssub:substituting_data}, it is now the relative entropy between two separately derived posteriors to be analyzed in the Gaussian limit. For simplicity, priors are furthermore set to be very wide compared to posterior constraints, such that the posteriors for $\mathcal L_{old} = \mathcal N(\mathcal D_{old};F(\Theta),\mathcal C_{old})$, $\mathcal L_{new} = \mathcal N(\mathcal D_{new};F(\Theta),\mathcal C_{new})$, and a linear model are given by
\begin{equation}
	p_{i}(\Theta) \approx \mathcal N (\Theta; \Theta_i, \Sigma_i),
\end{equation}
with
\begin{align}
	\Sigma_i = (M^T \mathcal C_i^{-1} M)^{-1},\\
	\Theta_i = \Sigma_i M^T \mathcal C_i(\mathcal D_i-F_0),
\end{align}
where $i$ is either $old$ or $new$. Note that in this case, $\Sigma_i$ is exactly given by the Fisher matrix associated to $\mathcal L_i$. Evaluating $D(p_{new}||p_{old})$ using equation \eqref{eq:ggre} is straightforward:
\begin{equation}
	\begin{aligned}
		&D(p_{new}||p_{old}) = \frac 1 2 (\Theta_{old} - \Theta_{new})^T\Sigma_{old}^{-1}(\Theta_{old} - \Theta_{new})\\
		&\qquad+\frac 1 2 \left(\text{tr}\left(\Sigma_{old}^{-1}\Sigma_{new}\right) - d - \log \frac {\det \Sigma_{new}} {\det\Sigma_{old}}\right).
	\end{aligned}
	\label{eq:recd}
\end{equation}

Splitting the relative entropy in an expected and a surprise part can be done along similar lines as in section \ref{sub:adding_supplementary_data2}. The prior distribution of the relative entropy is induced by the probability of data $\mathcal D_{new}$ as it is expected from $p_{old}$:
\begin{equation}
	p(\mathcal D_{new}) = \int d\Theta\, p_{old}(\Theta)\mathcal L_{new}(\Theta;\mathcal D_{new}).
\end{equation}
As shown in Appendix \ref{sub:expected_relative_entropy}, this results in an expected relative entropy given by
\begin{equation}
	\ere =-\frac 1 2 \log \frac {\det \Sigma_{new}} {\det \Sigma_{old}} + \text{tr}(\Sigma_{old}^{-1}\Sigma_{new}).
	\label{eq:expre2}
\end{equation}
The standard deviation of the expected relative entropy $\sigma(D)$ is given by
\begin{equation}
	\begin{aligned}
		\sigma^2(D) = \frac 1 2 \text{tr}\left((\Sigma_{old}^{-1}\Sigma_{new} + \mathbb 1)^2\right).
	\end{aligned}
	\label{eq:sigmad2}
\end{equation}
Just as in section \ref{sub:adding_supplementary_data2}, the surprise $S$ is defined as the difference between observed and expected relative entropy, $S=D-\ere$, and is anticipated to be of order $\sigma(D)$. Details on how to numerically calculate the p-values of $S$ can again be found in Appendix \ref{sub:expected_relative_entropy}.

As already mentioned in section \ref{ssub:substituting_data}, there are also more complex analysis strategies to be considered when dealing with correlated datasets. While the example of a joint likelihood for two correlated sets of data is hard to model within the framework of this section, partial replacement of data actually amounts to considering a non-flat prior for the analysis of $\mathcal L_{new}$. The results for this more general case are presented in Appendix \ref{sub:expected_relative_entropy}.



\section{Application to CMB Experiments} 
\label{sec:information_gains_from_cmb}

\noindent The anisotropies in the CMB are a prediction of $\Lambda$CDM and inflationary models and have been measured with great precision. First detected in the 1960s, precision cosmology from CMB observations started with the Cosmic Background Explorer (COBE) launched in 1989 \cite{Wright:1996um} and has continued to the recently published Planck results \cite{Collaboration:2013ww,Collaboration:2013uv}. The observables that are of most cosmological interest in CMB observations are the power spectra of the temperature and polarization fluctuations on the sky.

Considering the measured CMB temperature anisotropies as an example, one wants to evaluate the power spectrum of $\frac {\delta T} T(\hat n)$, i.e. the deviation $\delta T$ from the average temperature $T$ in direction $\hat n$. This is done by expanding the temperature fluctuations in terms of spherical harmonics: $\frac {\delta T} T(\hat n) = \sum_{\ell = 1}^\infty \sum_{m = -\ell}^\ell a_{\ell m}Y_{\ell m}(\hat n)$. The multipoles $C_\ell$ of the temperature power spectrum can then be calculated via
\begin{equation}
	\langle a_{\ell m} a^*_{\ell'm'} \rangle = \delta_{\ell\ell'}\delta_{mm'}C_\ell,
	\label{eq:csubl}
\end{equation}
where the average is over all possible CMB realizations. In an experiment, $a_{\ell m}$ can be estimated from the measured map of temperature anisotropies. The estimator for the observed multipole $C^{obs}_\ell$ is given by $C^{obs}_\ell = \frac 1 {2\ell + 1} \sum_{-\ell\leq m\leq \ell} |a^{obs}_{\ell m}|^2$. $C^{obs}_\ell$ and has two sources of uncertainty: the measurement errors on the $a^{obs}_{\ell m}$s and the cosmic variance arising from the fact that only one CMB sky is observable.

Once the estimates for $C_\ell^{obs}$ and its errors are obtained, the next step is to construct a likelihood for the cosmological parameters. As the power spectrum depends on the cosmological parameters in a non-trivial way, the input of such a likelihood function is usually a numerically evaluated power spectrum $C_\ell^{mod}$ predicted by the parametrization of the cosmological model:
\begin{equation}
	\mathcal L(\Theta;C_\ell^{obs})=p(C_\ell^{obs}|C_\ell^{mod}(\Theta)).
\end{equation}
The likelihood as a function of the parameters $\Theta$ can hence only be evaluated numerically for a specific choice of parameters and is unknown in its analytic form.

Furthermore, the CMB likelihood is often a function of both cosmological parameters $\Theta$ and so called nuisance parameters $\nu$ that model the influence of instrumental or astronomical effects on the data:
\begin{equation}
	\mathcal L(\Theta, \nu) = p(\mathcal D|\Theta, \nu).
\end{equation}
In this case, the posterior is also a distribution on the joint parameter space of $\Theta$ and $\nu$: 
\begin{equation}
	\p(\Theta,\nu) = \frac {\mathcal L(\Theta, \nu) p(\Theta, \nu)} {\int d\Theta d\nu\, \mathcal L(\Theta, \nu) p(\Theta, \nu)},
\end{equation}
with $p(\Theta, \nu)$ being the prior on both cosmological and nuisance parameters. The quantity of interest, however, is the relative entropy between posterior and prior of the cosmological parameters only, so the dependence on the additional nuisance parameters has to be marginalized before estimating the relative entropy:
\begin{equation}
	\p(\Theta) = \int d\nu\, \p(\Theta,\nu).
\end{equation}

\subsection{Data} 
\label{sub:data}

\begin{figure*}[t]
\centering
\includegraphics[width=.7\linewidth]{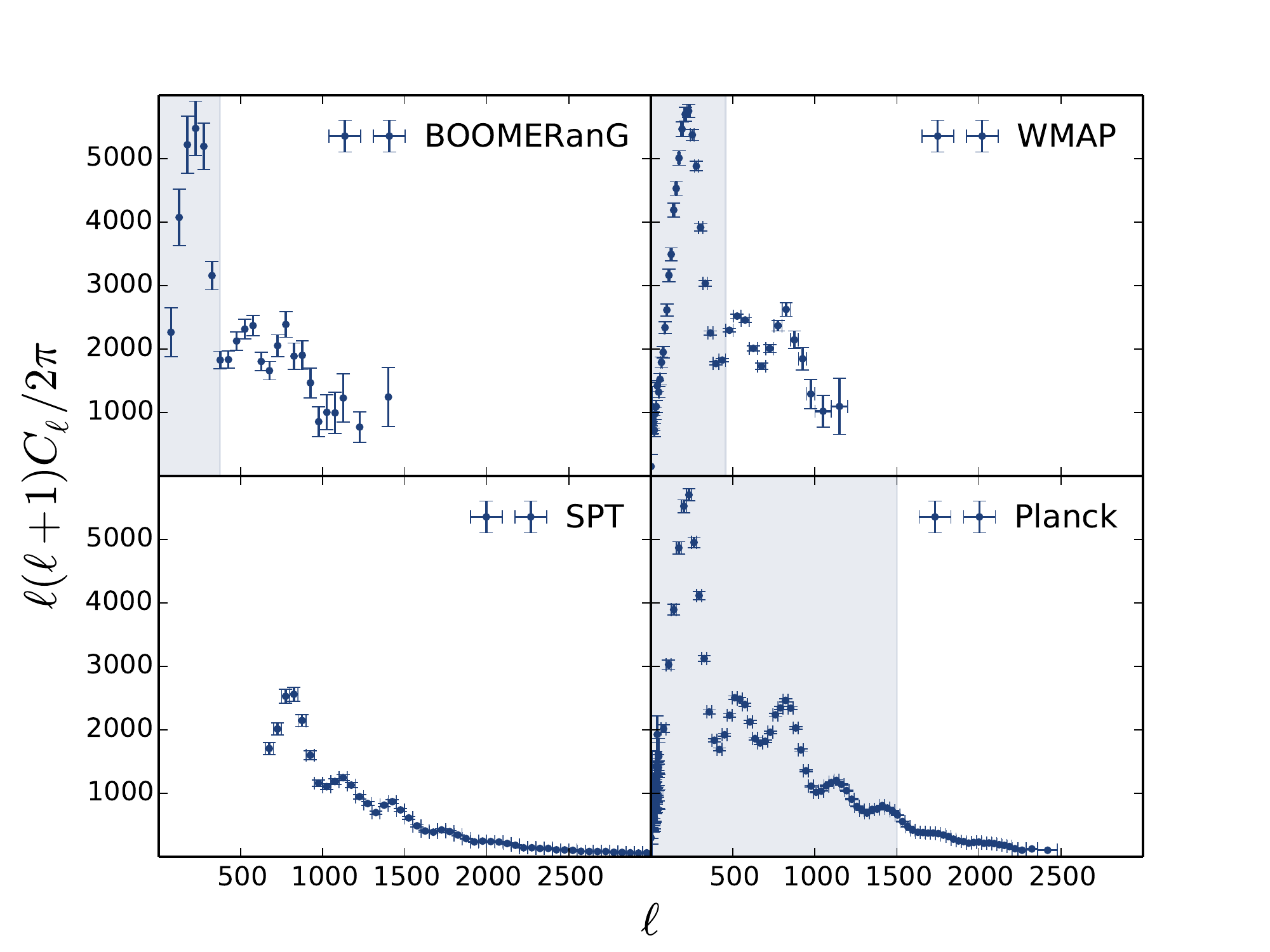}
\caption{Temperature power spectra from BOOMERANG, WMAP, SPT, and Planck. The shaded regions show the $\ell$-range in which the data is limited by cosmic variance.}
\label{fig:powerspecs}
\end{figure*}

\noindent In this work, the power spectra and likelihoods of four observations are considered: starting with the BOOMERANG data \cite{Jones:2006hd,Piacentini:2006wu,Montroy:2006vc}, the cosmological parameter constraints are updated with WMAP data \cite{Bennett:2012wy,Hinshaw:2012vg}, SPT data \cite{Story:2012vh}, and finally Planck data \cite{Collaboration:2013vc}. The temperature power spectra for each of these experiments are shown in Figure \ref{fig:powerspecs}. All four datasets are briefly discussed in the following.

\subsubsection{BOOMERANG} 
\label{ssub:boomerang}

The BOOMERANG data was collected during an antarctic balloon flight in 2003 \cite{Masi:2005vo}. The telescope which was attached to the balloon measured both temperature and polarization of the CMB over three sky regions of $90$, $750$, and $300$ square degrees and in three wide frequency bands centered at $145$, $245$, and $345$ GHz. The power spectra are estimated over the range $75 \leq \ell \leq 1400$ in temperature \cite{Jones:2006hd} and $150\leq \ell \leq 1000$ in polarization \cite{Piacentini:2006wu,Montroy:2006vc}. They are cosmic variance limited for $\ell<375$. The likelihood is a log-normal distribution in the temperature power spectrum and a normal distribution in polarization. It is numerically marginalized over a calibration factor and the size of the beam with an iterated Gauss-Legendre quadrature.


\subsubsection{WMAP} 
\label{ssub:wmap}

\noindent The all-sky survey of the CMB of the WMAP satellite had a duration of nine years, resulting in a measurement of both temperature and polarization power spectra of the CMB over the range $2 \leq \ell \leq 1200$ in temperature and $2 \leq \ell \leq 1000$ in polarization \cite{Bennett:2012wy,Hinshaw:2012vg}. The measurements are limited by cosmic variance for all $\ell < 457$. Together with the data, the WMAP team published a likelihood code which is used in this work without modifications. The data releases after three \cite{Spergel:2007ux}, five \cite{Dunkley:2009iz}, seven \cite{Larson:2011vd}, and nine years \cite{Bennett:2012wy,Hinshaw:2012vg} are considered.


\subsubsection{SPT} 
\label{ssub:spt}

\noindent The measurements of the ground based SPT survey \cite{Story:2012vh} focus on the small scale fluctuations of the CMB. Estimated from data of 2500 square degrees, the publicly available temperature power spectrum ranges from $650 \leq \ell \leq 3000$. As the CMB signal is contaminated by foreground effects on these scales, templates have to be removed from the observed power spectrum in order to model the foreground effects. This results in three parameters---accounting for the amplitude of the power from Poisson distributed point sources, clustered point sources, and Sundaev-Zeldovich clusters---that additionally enter the Gaussian likelihood for the cosmological parameters. As these parameters come with Gaussian priors, they can, however, be marginalized theoretically (see Appendix \ref{sec:marginalizing_template_amplitudes} for more details).


\subsubsection{Planck} 
\label{ssub:planck}

\noindent The final data is from the all-sky measurements of the CMB temperature by the Planck satellite \cite{Collaboration:2013ww}. The power spectrum as observed by Planck covers the range $2 \leq \ell \leq 2500$. The likelihood function is split into a low-$\ell$ part and a high-$\ell$ at $\ell=50$. While the low-$\ell$ part depends on the cosmological parameters, the high-$\ell$ likelihood has to model foregrounds just as in the SPT likelihood. The Planck team uses $16$ additional nuisance parameters with mostly flat priors to describe the foregrounds in great detail.



\subsection{Estimation of relative entropy} 
\label{sub:estimation_of_relative_entropy}

\noindent The CMB likelihood is usually not known analytically but has to be evaluated numerically. Consequently, the relative entropy between two posteriors has to be estimated numerically, too. The two Monte Carlo techniques that were used for the estimation are quickly outlined here but explained in more detail in Appendix \ref{sec:relative_entropy_estimation_with_monte_carlo_methods}.

The first procedure assumes Gaussianity of the underlying distributions. Using the analytic expression from section \ref{sec:the_gaussian_limit} for the relative entropy between normal distributions, it is estimated from the moments of posterior samples generated with the Monte Carlo Markov chain framework \verb|CosmoHammer| by \citet{Akeret:2012uk}. \verb|CosmoHammer| uses \verb|emcee| \cite{ForemanMackey:2012ux} as Monte Carlo Markov chain algorithm, an implementation of the affine-invariant samplers proposed by \citet{Goodman:2010eu}. \verb|CAMB| \cite{Lewis:2008wx} is used for calculating the theoretical power spectra. The $\Lambda$CDM model is parametrized with today's Hubble parameter $H_0$, baryon density $\Omega_bh^2$, and dark matter density $\Omega_ch^2$. In addition, the optical depth due to reionization $\tau$ and the power law index $n_s$ and amplitude $A_s$ of the primordial curvature power spectrum are used. Within this framework, the marginalization of additional nuisance parameters is straightforward. Whenever possible, the public chains from the WMAP and Planck teams served as a consistency check for the findings.

In the second procedure, one effectively performs a Monte Carlo integration to estimate the relative entropy (see Appendix \ref{sub:general_approach} for more details). It is not restricted to Gaussian distributions but requires knowledge of the likelihood as a function of the cosmological parameters only. As the Planck likelihood is a joint function of both nuisance and cosmological parameters, relative entropies involving the Planck likelihood are restricted to the Gaussian approximation.


\subsection{Numerical results} 
\label{sub:numerical_results}

\noindent Because the CMB power spectra are correlated due to cosmic variance, only datasets that have a small overlap in the measured scale of the power spectra or temperature and polarization datasets can be combined in a simple sequential analysis. Examples for both joint analyses of complementary data and separate analyses of correlated data are discussed in the following, considering combinations of the datasets introduced in section \ref{sub:data}.

\begin{table*}[t]
	\centering
	\caption{Numerical relative entropy estimates in bits for considered combinations of CMB data. For the Gaussian approximation, the relative entropy $D$ is split into expected relative entropy $\ere$ and surprise $S = D- \ere$. Furthermore, the expected spread $\sigma(D)$ of $D$ around its mean $\ere$ and the significance of the surprise $S/\sigma(D)$ are given. Depending on the analysis strategy, $\ere$ and $\sigma(D)$ are given by \eqref{eq:expre} and \eqref{eq:sigmad} when adding data, by \eqref{eq:expre2} and \eqref{eq:sigmad2} when replacing data, and by \eqref{eq:generalere} and \eqref{eq:sigmadgen} for partial replacement. For joint analyses, $\ere$ and $\sigma(D)$ are calculated as if the data was added independently. The p-value is an estimate for the prior probability for observing a surprise that is greater or equal (less or equal) than $S$ if $S$ is greater (smaller) than zero. It is an approximation when data is partially replaced.}
	\begin{ruledtabular}
	\begin{tabular}{rclccccccc}
		\multicolumn{3}{c}{Data combination\footnote{WMAP 9 = full WMAP 9 data (same for WMAP 3, 5, and 7), WP = WMAP 9 polarization data, Planck = Planck temperature data, SPT = SPT temperature data, BOOMERANG = full BOOMERANG data}} & Updating & \multicolumn{5}{c}{Gaussian approximation\footnote{The errors of the Gaussian estimates for $D$, $\ere$, $S$, and $\sigma(D)$ are of order $0.1$.}} & Monte Carlo\\
		&&& scheme\footnote{add = adding data, replace = replacing data, part = partial replacement of data, joint = joint analysis of data} & $D$ & $\ere$ & $S$ & $S / \sigma(D)$ & p-value\footnote{See appendix \ref{sub:expected_relative_entropy} for details on the estimation of the p-value.} & estimate of $D$\footnote{The results from the Monte Carlo integration are stated including the estimation uncertainty.} \\\hline
		BOOMERANG &$\rightarrow$& WMAP 9 & replace & $22.5$ & $18.4$ & $4.1$ & $1.6$ & $0.07$ &$20.9 \pm 0.6$\\
		WMAP 3 &$\rightarrow$& WMAP 5 & joint & $7.7$ & $2.2$ & $5.5$ & $5.3$ & $0.001$ & $10.5 \pm 0.9$\\
		WMAP 5 &$\rightarrow$& WMAP 7 & joint & $1.4$ & $1.0$ & $0.4$ & $0.6$ & $0.2$ & $1.5 \pm 0.7$\\
		WMAP 7 &$\rightarrow$& WMAP 9 & joint & $1.5$ & $1.2$ & $0.3$ & $0.4$ & $0.3$ & $1.3 \pm 0.7$\\
		WMAP 9 &$\rightarrow$& WMAP 9 + SPT & add & $4.3$ & $2.1$ & $2.2$ & $2.1$ & $0.04$ &$4.6\pm0.7$\\
		WMAP 9 &$\rightarrow$& Planck + WP & part & $29.8$ & $7.9$ & $21.9$ & $6.5$ & $0.0002$ & ---\\
		WMAP 9 + SPT &$\rightarrow$& Planck + WP + SPT & part & $27.8$ & $6.6$ & $21.2$ & $6.5$ & $0.0002$ & ---\\
		Planck &$\rightarrow$& Planck + WP & add & $1.2$ & $2.2$ & $-0.9$ & $-0.9$ & $0.08$ & ---\\
	\end{tabular}
	\end{ruledtabular}
	\label{tab:relent}
\end{table*}

\begin{figure*}[p]
	\centering
	\includegraphics[width=.53\linewidth]{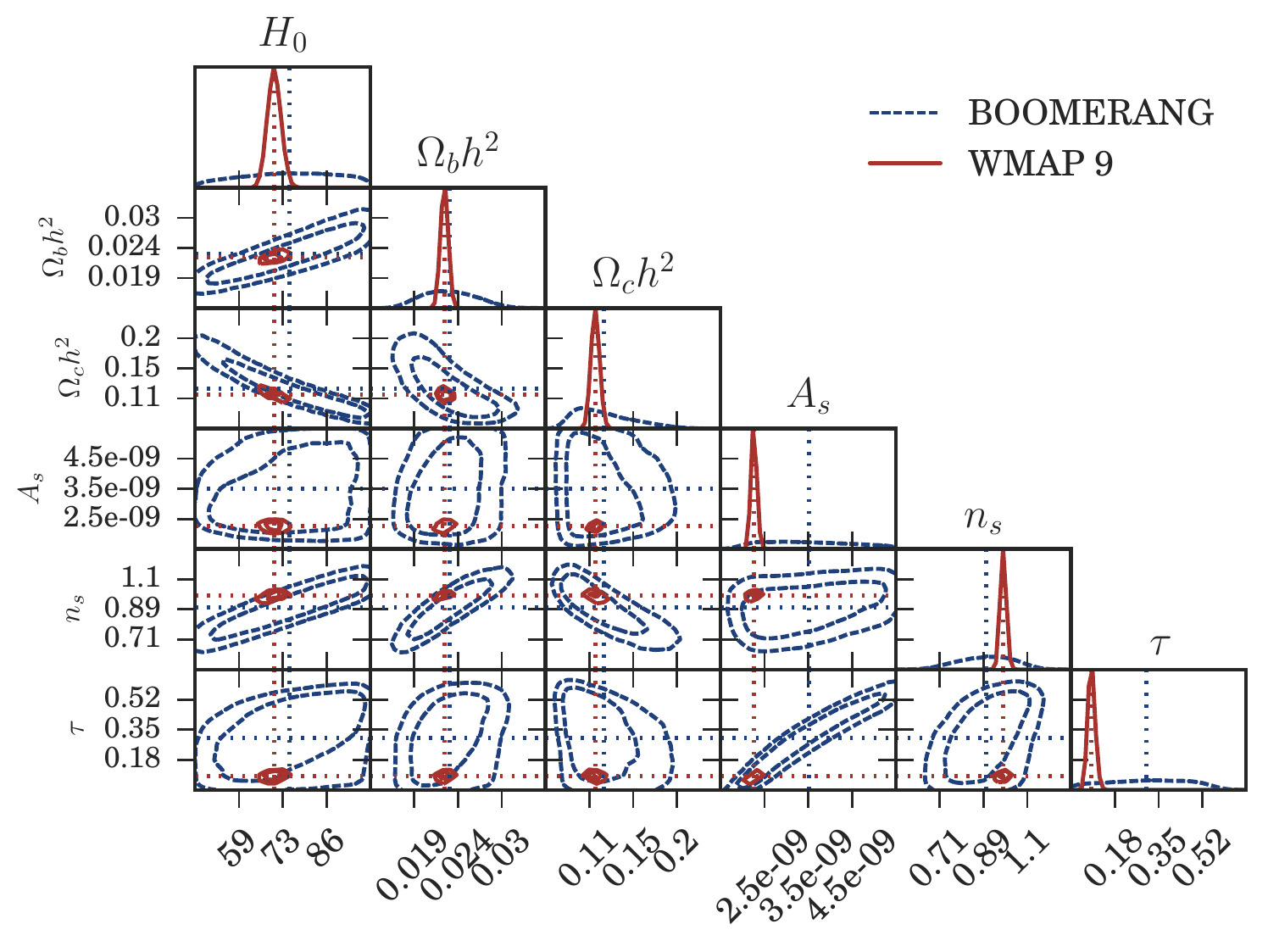}
	\caption{Marginalized posteriors from BOOMERANG and WMAP 9 data. Dotted lines show the means.}
	\label{fig:boommarginals}
	\includegraphics[width=.53\linewidth]{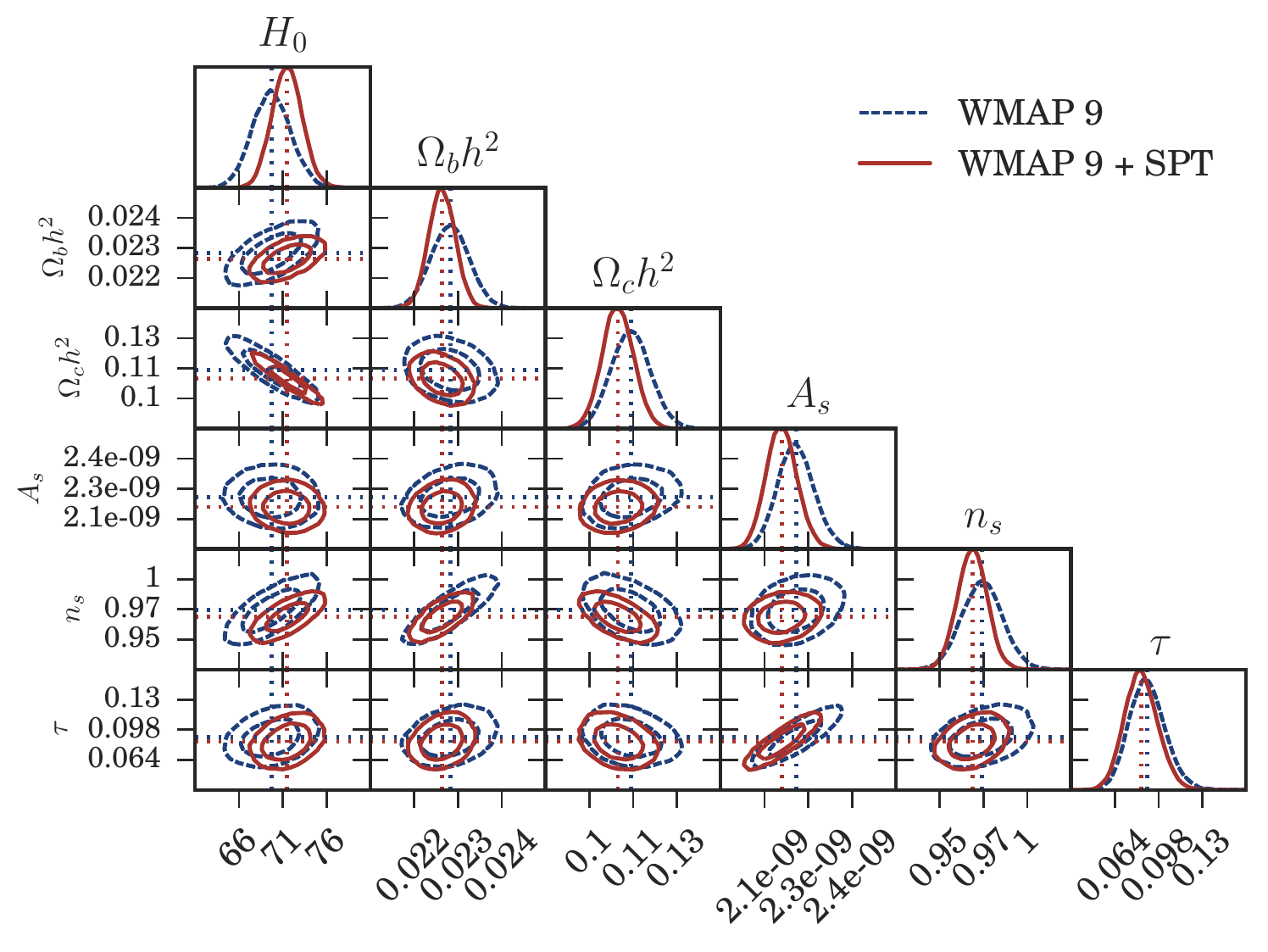}
	\caption{Marginalized distributions of the WMAP 9 constraints with and without SPT data. Dotted lines show the means.}
	\label{fig:sptmarginals}
	\includegraphics[width=.53\linewidth]{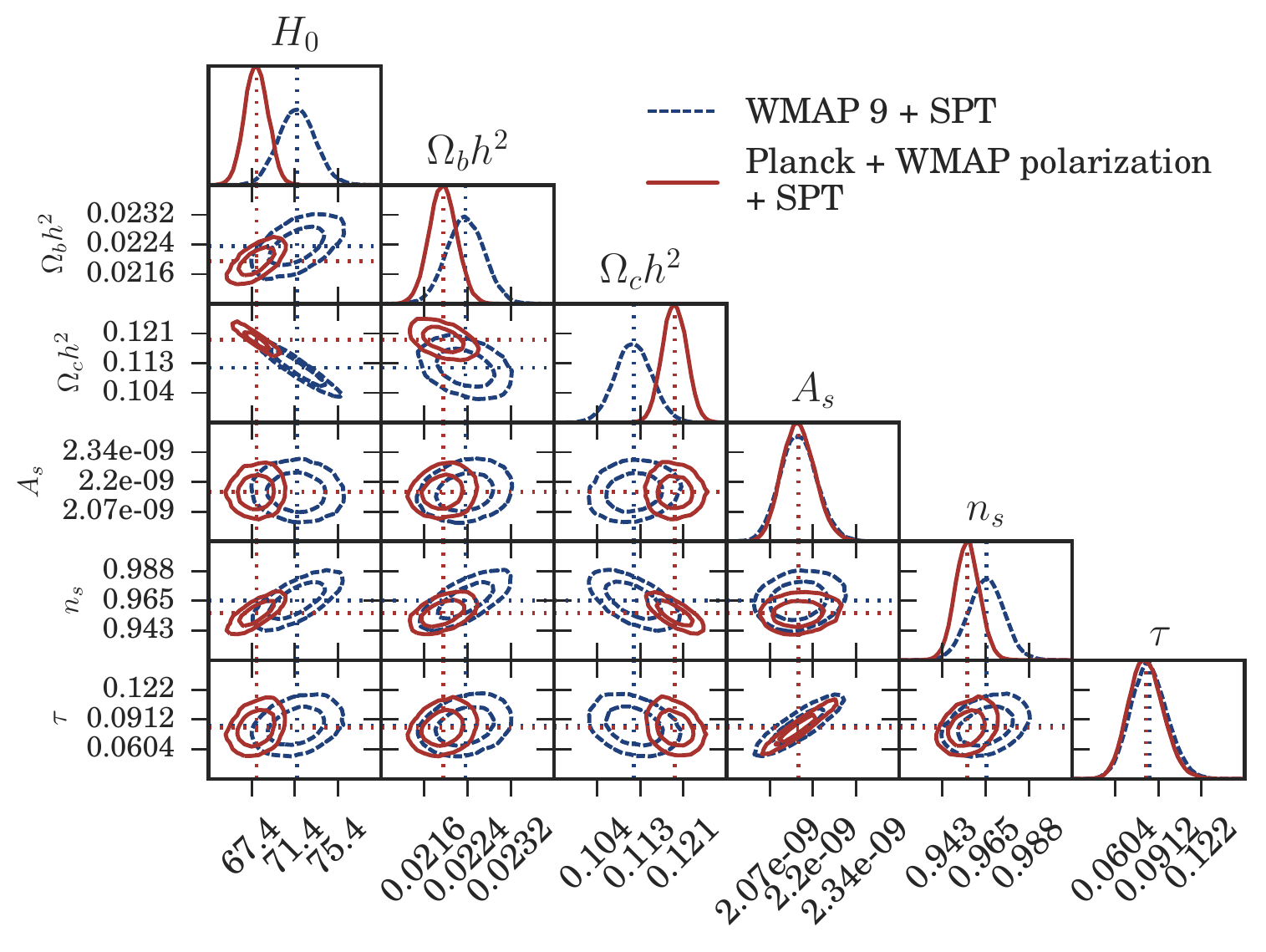}
	\caption{Marginalized posteriors from WMAP 9 and Planck when combined with SPT data. Dotted lines show the means.}
	\label{fig:wmapvsplanckmarginals2}
\end{figure*}

\subsubsection{Replacing BOOMERANG with WMAP data} 
\label{ssub:replacing_boomerang_data_with_wmap}

\noindent As can be seen in Figure \ref{fig:powerspecs}, BOOMERANG data overlaps completely with WMAP data. Furthermore, WMAP observations are more accurate than the measurements from BOOMERANG, so a full joint analysis would provide only modest improvements compared to simply replacing the BOOMERANG with WMAP observations. Comparing BOOMERANG and WMAP data is therefore an example of datasets that replace each other as discussed in section \ref{ssub:substituting_data}. The separately analyzed posteriors of the two experiments are shown in Figure \ref{fig:boommarginals}. It can be seen that while $\Omega_bh^2$, $\Omega_ch^2$ and $n_s$ are reasonably well constrained, $H_0$ is almost unconstrained by BOOMERANG data and $A_s$ and $\tau$ are highly degenerate. Nevertheless, the relative entropy estimates from the Gaussian approximation and the Monte Carlo method agree well as can be seen in Table \ref{tab:relent}. The total gain from this replacement is $22.5$ bits. This significant update can also be seen in Figure \ref{fig:boommarginals}, which shows that most of this gain can be attributed to a drastic reduction in the volume of the confidence intervals. This is further demonstrated by the decomposition of the relative entropy in Table \ref{tab:relent}, where the dominant contribution ($18.4$ bits) comes from $\ere$, the expected entropy gain. The surprise from this update is smaller than $2\sigma(D)$. The fact that the WMAP likelihood is not strictly of the Gaussian type given in equation \eqref{eq:normlike} and used in section \ref{sec:the_gaussian_limit} implies that $\ere$ and $\sigma(D)$ are approximations. Nevertheless, the conclusions drawn from Table \ref{tab:relent} and Figure \ref{fig:boommarginals} are apparently consistent.


\subsubsection{Comparing the individual WMAP releases} 
\label{ssub:comparing_the_wmap_releases}

\noindent The relative entropy gains for the WMAP data releases after collecting three, five, seven, and nine years of data are shown in Table \ref{tab:relent}. As the data of the individual years is correlated, the WMAP team published joint likelihood functions for the overall data, taking those correlations into account. A comparison of the joint analyses of the WMAP data is therefore a mixture of the updating types discussed in section \ref{sub:relative_entropy}, but best described by a sequential updating of the constraints from old data with additional new data. For the estimates of $\ere$, $S$, and $\sigma(D)$ shown in Table \ref{tab:relent}, it was therefore assumed that the data is complimentary, as a joint analysis cannot be modeled in the framework of section \ref{sec:the_gaussian_limit}. The WMAP 3 posterior weakly deviates from Gaussianity, while the other posteriors are well described by a multivariate normal distribution. This deviation from Gaussianity leads to the difference between the relative entropy estimates using Gaussian approximations and the Monte Carlo integration method. The update from WMAP 3 to WMAP 5 shows a significantly larger increase in relative entropy ($7.7$ bits) as compared to the updates to WMAP 7 and 9 ($1.4$ and $1.5$ bits, respectively). However, on closer inspection it can be seen that the majority of this gain is in the surprise ($5.5$ bits) which is at the $5\sigma(D)$ level, corresponding to a p-value of $0.001$ as the distribution for $D$ is non-Gaussian. This unexpectedly strong change from the three to the five year release can be partly attributed to a change in the likelihood for the low-$\ell$ temperature power spectrum \cite{Dunkley:2009iz}.

\subsubsection{Updating WMAP constraints with SPT data} 
\label{ssub:updating_wmap_constraints_with_spt_data}

As can be seen in Figure \ref{fig:powerspecs}, SPT and WMAP data have little overlap in the high-$\ell$ regime where correlations due to cosmic variance are small, and can hence be considered as complementary datasets. As discussed in section \ref{ssub:adding_supplementary_data}, the constraints from WMAP are therefore compared to the posterior after adding the SPT data to the WMAP constraints. The marginals of prior and posterior when analyzing the SPT data with a WMAP 9 prior are shown in Figure \ref{fig:sptmarginals}. The estimates for the relative entropy are listed in Table \ref{tab:relent}. Since both distributions are well approximated by multivariate Gaussians, the Gaussian approximation yields reliable results as can be seen by comparing it to the Monte Carlo integration. Furthermore, the SPT likelihood is a normal distribution in the data. As such, the requirements from section \ref{sec:the_gaussian_limit} are fulfilled to good approximation and splitting $D$ into $\ere$ and $S$ according to equations \eqref{eq:expre} and \eqref{eq:surprise} is justified. The information gain here is $4.3$ bits with $2.1$ bits coming from $\ere$, which is comparable to the update from WMAP 5 to WMAP 9, and a surprise at the $2\sigma(D)$ level.


\subsubsection{Impact of Planck}
\label{ssub:impactofplanck}

\noindent WMAP temperature data and Planck observations are strongly correlated particularly on large scales. Hence, in the analysis shown in Table \ref{tab:relent} the WMAP data is partially replaced by the temperature data from Planck while WMAP polarization (WP) data (with and without SPT data) is used in both analyses. When Planck is added to previous data (with and without SPT) there are large gains in relative entropy ($29.8$ and $27.8$ bits). When studying the decomposition, however, it can be seen that the contribution from $\ere$ to the total entropy gain is dominated by the surprise part ($21.9$ and $21.2$ bits), though measuring a considerable improvement in precision at $7.9$ and $6.6$ bits. Furthermore, the surprise is at levels greater than $6\sigma(D)$ corresponding to a p-value of $0.0002$. The results shown in Figure \ref{fig:wmapvsplanckmarginals2} support these findings and show that though the error contours do decrease considerably with the addition of Planck constraints another apparent effect is the shift of the confidence intervals. This in fact echoes the results of the \citet{Collaboration:2013uv} demonstrating shifts of the order of a standard deviation in four parameters when comparing WMAP 9 constraints to the ones from Planck and WMAP polarization data. The relative entropy analysis of the posteriors is hence able to detect inconsistencies between posteriors in the case when none of the shifts in the individual parameters is particularly significant on its own while the overall shift in parameter space is nevertheless significantly larger than expected.

To further study the origin of this surprise contribution, it is illustrative to estimate the relative entropies when replacing large scales ($2 \leq \ell \leq 49$) and small scales ($\ell \geq 50$) of the temperature power spectrum separately. Discrepancies between Planck and WMAP have been found on large scales when taking cosmic variance correlations into account \cite{Collaboration:2013vc}. However, these discrepancies only have a small effect on the cosmological parameters: The estimated relative entropy of $0.7$ bits when replacing large scale data only has a small surprise contribution of $-0.1$. Note that the negative surprise is due to the correlations between Planck and WMAP data in the low-$\ell$ regime. The overall surprise is therefore not caused by large scales. Instead, one finds $D = 29.2$ and a surprise of $21.4$ when only small scale data is replaced.

Table \ref{tab:relent} also shows the effect on the parameters when adding WMAP polarization data to the Planck measurements. The findings show that WMAP polarization data adds $1.2$ bits and that the surprise is negative, i.e. the means shift less than expected.




\section{Conclusions} 
\label{sec:discussion}

\noindent In order to compare the cosmological parameter constraints from different experiments, a tool for quantifying changes in posterior distributions on the full parameter space is needed. Motivated from information theory, the concept of relative entropy measures differences between distributions in a parametrization independent way and is therefore able to quantify the information gained from new data. In this work, relative entropy is used to develop a new tool for comparing the parameter constraints of the $\Lambda$CDM model from different CMB surveys. Two ways of combining data from different experiments are discussed: complementary datasets that can be analyzed sequentially and correlated measurements that replace earlier datasets.

Relative entropy captures both changes in confidence volumes and location of the regions of the posteriors. In the regime of Gaussian likelihoods and linear models, these contributions can even be distinguished as an expected relative entropy measuring differences in confidence volume and a surprise coming from shifts in parameter space. This Gaussian regime is furthermore at least a good approximation for CMB data analysis. The notions of expected relative entropy and surprise turn the relative entropy into a powerful diagnostic for the consistency of datasets.

The relative entropy gains in units of bits from BOOMERANG, WMAP, SPT, and Planck surveys range from about 1 to 30. In general, the numbers are driven by the contributions from the expected relative entropy, $\ere$, but in three cases the surprise is found to dominate the results. In terms of expected relative entropy, the step from Boomerang to WMAP is the biggest ($\ere \sim 18$ bits), followed by the update of WMAP by Planck data ($\ere \sim 7$ bits). The addition of SPT data to the WMAP constraints leads to an expected relative entropy gain of $\ere \sim 2$ bits.

Looking at the total relative entropy gains, inclusion of Planck data shows the biggest gains ($D=29.8$ bits from WMAP 9 and $D=27.8$ bits from WMAP 9 and SPT). When these numbers are decomposed, the relative entropy is found to be dominated by the surprise ($S=21.9$ and $S=21.2$ bits, respectively). These are very significant surprise values since they are $6.5$ standard deviations from expectations. Note that the expected distribution of $D$ is non-Gaussian; for the corresponding p-values see Table \ref{tab:relent}. This indicates that the shifts in the confidence intervals of the posteriors are large compared to the shifts expected from the increased precision. This conclusion is further supported by the changes in the marginalized posterior plots and points to possible tensions when fitting predictions from $\Lambda$CDM to different measurements, in line with other findings \cite{Verde:2013hp,Spergel:2013va}. Other updates considered here also show significant surprise. In particular the update from WMAP 3 to WMAP 5 shows a large surprise of $S=5.5$ which is $5.3$ standard deviations away from expectations. This result might be caused by both the deviations of the WMAP 3 posterior from a normal distribution and the adjustments of the likelihood function for the low-$\ell$ temperature power spectrum by the WMAP team. It is also interesting to note that while it is possible and expected for the surprise to be both positive and negative, the findings presented here typically show positive surprise. The reason for this is unclear, but if measurement errors are systematically underestimated this in itself would tend to bias the results towards positive surprise.

To conclude, the relative entropy is found to be a valuable diagnostic to compare constraints from different measurements. In cases where the likelihood is close to Gaussian and the model is effectively linear, the contributions from shifts in the confidence regions can be separated from the gains in precision. The resulting quantities are easy to estimate and are capable of describing the overall changes in multidimensional constraints in an efficient way.


\section*{Acknowledgements} 
\label{sec:acknowledgements}

\noindent This work was in part supported by the Swiss National Science Foundation (grant number $200021\_143906$). We thank the anonymous referee for useful comments on the manuscript.


\appendix

\section{Relative Entropy and Normal Distributions} 
\label{sec:relative_entropy_and_normal_distributions}

\noindent The statements on relative entropy and normal distributions introduced in section \ref{sec:the_gaussian_limit} are derived next.

\subsection{Deriving the posterior} 
\label{sub:deriving_the_posterior}

\noindent Here, the posterior given in equation \eqref{eq:ggpost} is derived, where likelihood, prior and model are given by
\begin{align}
	p(\Theta) &= \mathcal N (\Theta; \Theta_p, \Sigma_p),\\
	\mathcal L(\Theta; \mathcal D) &= \mathcal N(\mathcal D; F(\Theta), \mathcal C),\\
	F(\Theta) &= F_0 + M\Theta,
\end{align}
with $\Theta \in \mathbb R^d$ and $\mathcal D \in \mathbb R^n$. The posterior is then defined by equation \eqref{eq:bayest}. To show that the posterior is normally distributed it is useful to define $\delta \mathcal D = \mathcal D - F_0\in\mathbb R^n$, the linear subspace $W\subset\mathbb R^n$:
\begin{equation}
	W \equiv \{O\in\mathbb R^n:\exists\Theta\in\mathbb R^d \text{ s.th. } O = M\Theta\},
\end{equation}
and its orthogonal complement with the respect to the bilinear form $B(x,y) \equiv x^T\mathcal C^{-1}y$:
\begin{equation}
	W^\perp \equiv \{O\in\mathbb R^n:B(O,O')=0\;\forall O'\in W\}.
\end{equation}
One can now decompose $\delta \mathcal D$ as $\delta \mathcal D = \delta \mathcal D^\perp + \delta \mathcal D^\parallel$ with $\delta \mathcal D^\perp\in W^\perp$ and $\delta \mathcal D^\parallel\in W$. As $\delta \mathcal D^\parallel\in W$, there exists a $\Theta_\mathcal L$ such that $\delta \mathcal D^\parallel = M\Theta_\mathcal L$. One can hence rewrite the likelihood as follows:
\begin{equation}
	\begin{aligned}
		-2\log \mathcal L(\Theta;\mathcal D) &\propto (\delta \mathcal D - M\Theta)^T C^{-1} (\delta \mathcal D - M\Theta)\\
		&= (\Theta_\mathcal L - \Theta)^T M^T\mathcal C^{-1}M(\Theta_\mathcal L - \Theta)\\
		&\qquad +(\delta \mathcal D^\perp)^TC^{-1}(\delta \mathcal D^\perp)\\
		&\propto \mathcal N(\Theta; \Theta_\mathcal L, (M^T\mathcal C^{-1}M)^{-1}),
	\end{aligned}
	\label{eq:likenorm}
\end{equation}
showing that the likelihood is indeed proportional to a Gaussian in $\Theta$ with mean $\Theta_\mathcal L$ and covariance matrix $(M^T\mathcal C^{-1}M)^{-1}$. As $\p(\Theta) \propto \mathcal L(\Theta) p(\Theta)$, it is straightforward to show that the posterior is a normal distribution. Using that the Fourier transform of a Gaussian is of the form
\begin{equation}
	\mathcal N(X; \mu, \Sigma) = \int \frac {d^dk} {(2\pi)^d} e^{-ik^T(X-\mu)} e^{\frac 1 2 k^T\Sigma k},
\end{equation}
it is easy to calculate the evidence $E(\mathcal D)$ as well as mean $\Theta_{new}$ and the covariance matrix $\Sigma_{new}$ of the posterior:
\begin{align}
	E(\mathcal D) &\equiv \int d\Theta\, \mathcal L(\Theta;\mathcal D)p(\Theta)\nonumber\\
	&=\mathcal N(\mathcal D; F(\Theta_p), \mathcal C + M\Sigma_pM^T),\\
	\Theta_{new} &\equiv \int d\Theta\, \Theta\mathcal L(\Theta;\mathcal D)p(\Theta)\nonumber\\
	&= \Sigma_{new} \left(\Sigma_p^{-1}\Theta_p + M^T\mathcal C^{-1}(\mathcal D - F_0) \right),\\
	(\Sigma_{new})_{ij} &\equiv \int d\Theta\, \Theta_i\Theta_j\mathcal L(\Theta;\mathcal D)p(\Theta) - \Theta_{new, i}\Theta_{new, j}\nonumber\\
	&= \left(\Sigma_p^{-1} + M^T\mathcal C^{-1}M\right)^{-1}_{ij}.
\end{align}

\subsection{Relative entropy of two gaussians} 
\label{sub:relative_entropy_between_gaussians}

\noindent Let $P_1(X) = \mathcal N(X, X_1, \Sigma_1)$ and $P_2(X) = \mathcal N(X, X_2, \Sigma_2)$ with $X$ being $d$ dimensional and define $\Delta X_i = X - X_i$. The relative entropy between $P_1$ and $P_2$ is then given by
\begin{equation}
	\begin{aligned}
		&D(P_1||P_2) = \int dX\, P_1(X) \log \frac {P_1(X)} {P_2(X)}\\
		&= -\frac 1 2 \left( \langle \Delta X_1^T\Sigma_1^{-1}\Delta X_1 \rangle_{P_1}\right.\\
		&\qquad\left. - \langle \Delta X_2^T\Sigma_2^{-1}\Delta X_2 \rangle_{P_1} - \log \frac {\det(\Sigma_1)} {\det(\Sigma_2)}\right) \\
		&= \frac 1 2 \left(-\text{tr}(\Sigma_1\Sigma_1^{-1}) + (X_1 - X_2)^T\Sigma_2^{-1}(X_1 - X_2) \right.\\
		&\qquad\left. + \langle \Delta X_1^T\Sigma_2^{-1}\Delta X_1\rangle_{P_1} - \log \frac {\det(\Sigma_1)} {\det(\Sigma_2)} \right)\\
		&= 	\frac 1 2 (X_1 - X_2)^T \Sigma_2^{-1} (X_1 - X_2)\\
		&\qquad +\frac 1 2\left(\text{tr}(\Sigma_1\Sigma_2^{-1}) - d - \log \frac {\det(\Sigma_1)} {\det(\Sigma_2)} \right).
	\end{aligned}
\end{equation}

\subsection{Distribution of relative entropy} 
\label{sub:expected_relative_entropy}

\noindent In this Appendix, the distribution of the relative entropy between two posteriors is discussed. In the most general case considered, there are three distributions of interest: the posterior of the old observation $p_1(\Theta)$, the prior $q(\Theta)$ for the analysis of a second observation, and the likelihood $\mathcal L(\Theta;\mathcal D)$ of the second observation. In this appendix, all of these distributions are considered to be Gaussian:
\begin{align}
	p_1(\Theta) = \mathcal N(\Theta;\Theta_1,\Sigma_1),\\
	q(\Theta) = \mathcal N(\Theta;\Theta_q,\Sigma_q),\\
	\mathcal L(\Theta;\mathcal D) = \mathcal N(\mathcal D;F(\Theta),\mathcal C),
\end{align}
while the model is chosen to be linear in $\Theta$:
\begin{equation}
	F(\Theta)=F_0 + M\Theta.
\end{equation}
According to Appendix \ref{sub:deriving_the_posterior}, the posterior $p_2$ derived from prior $q$ and likelihood $\mathcal L$ is then given by
\begin{equation}
	p_2(\Theta) = \mathcal N(\Theta;\Theta_2,\Sigma_2),
\end{equation}
with
\begin{align}
	\Sigma_2 &= (\Sigma_q^{-1} + M^T\mathcal C^{-1}M)^{-1},\\
	\Theta_2 &= \Sigma_2(\Sigma_q^{-1}\Theta_q + M^TC^{-1}(\mathcal D-F_0)).\label{eq:theta2}
\end{align}
Using the result from Appendix \ref{sub:relative_entropy_between_gaussians}, it is straightforward to calculate the relative entropy between the posteriors:
\begin{equation}
	\begin{aligned}
		D(p_2||p_1) &= \frac 1 2 \left( \text{tr}\left(\Sigma_1^{-1}\Sigma_2\right) - d - \log \frac {\det \Sigma_2} {\det\Sigma_1} \right.\\
		&\qquad+\underbrace{\left.(\Theta_2 - \Theta_1)^T\Sigma_1^{-1}(\Theta_2 - \Theta_1)\right)}_{\equiv\Delta(\mathcal D)},
	\end{aligned}
	\label{eq:apprelprpo}
\end{equation}
where the second line depends on the data $\mathcal D$ via $\Theta_2$ from equation \eqref{eq:theta2}. The prior distribution for $\mathcal D$ derived from the old posterior $p_1$ and the likelihood $\mathcal L_2$ is calculated in section \ref{sub:deriving_the_posterior} and given by
\begin{equation}
	\begin{aligned}
		p(\mathcal D) &= \int d\Theta\, \mathcal L(\Theta;\mathcal D)p_1(\Theta)\\ 
		&= \mathcal N(\mathcal D;F(\Theta_1),\mathcal C+M\Sigma_1M^T).
	\end{aligned}
	\label{eq:pev}
\end{equation}
As only the $\Delta$ part of equation \eqref{eq:apprelprpo} depends on $\mathcal D$, the prior distribution of $D(p_2||p_1)$ is equivalent to the distribution of $\Delta$ up to a shift. Focusing on this term one finds for $\delta\Theta\equiv\Theta_2 - \Theta_1$:
\begin{equation}
	\begin{aligned}
		\delta\Theta &= \Sigma_2 \left(\Sigma_q^{-1}\Theta_q + M^T\mathcal C^{-1}(\mathcal D - F_0) \right) - \Theta_1\\
		&=\Sigma_2\left(M^T\mathcal C^{-1}\left(\mathcal D - F(\Theta_1)\right) + \Sigma_q^{-1}(\Theta_q - \Theta_1)\right)\\
		&=\Sigma_2M^T\mathcal C^{-1}\left(\mathcal D - F(\Theta_1) + MQ^{-1}T\right),
	\end{aligned}
\end{equation}
with
\begin{align}
	Q &\equiv M^T\mathcal C^{-1} M = \Sigma_2^{-1} - \Sigma_q^{-1},\\
	T &\equiv \Sigma_q^{-1}(\Theta_q - \Theta_1).
\end{align}
Plugging $\delta\Theta$ into $\Delta$, one finds
\begin{equation}
	\Delta = \delta\Theta \Sigma_1^{-1}\delta\Theta = X^TAX,
	\label{eq:genchi2}
\end{equation}
with
\begin{align}
	X&\equiv \mathcal D - F(\Theta_1) + MQ^{-1}T,\\
	A&\equiv \mathcal C^{-1}MWM^T\mathcal C^{-1},\label{eq:defA}\\
	W&\equiv \Sigma_2\Sigma_1^{-1}\Sigma_2.
\end{align}
As $\mathcal D$ is distributed as a normal distribution with mean $F(\Theta_1)$ and covariance matrix $\mathcal C+M\Sigma_1M^T$, the new variable $X$ is also distributed as a normal distribution with mean $\mu$ and covariance matrix $\Sigma$ now given by
\begin{align}
	\mu &= MQ^{-1}T,\label{eq:defmu}\\
	\Sigma &= \mathcal C+M\Sigma_1M^T.\label{eq:defSigma}
\end{align}
The distribution of $\Delta$ is therefore the distribution of a quadratic form in $X$ where $X$ is normally distributed and is usually called a generalized $\chi^2$ distribution. Using textbook results for generalized $\chi^2$ distributions, the moments of $\Delta$ and hence of $D(p_2||p_1)$ can be easily derived. According to chapter 3.2b of \cite{mathai1992quadratic}, mean and variance of $\Delta$ are given by
\begin{align}
	\text{E}(\Delta)&=\text{tr}(A\Sigma) + \mu^TA\mu,\\
	\text{Var}(\Delta)&=2\text{tr}\left((A\Sigma)^2\right) + 4 \mu^TA\Sigma A\mu,
\end{align}
where $\text{E}(\Delta)$ is the expectation value of $\Delta$ and $\text{Var}(\Delta)$ is its variance. Using definitions \eqref{eq:defA}, \eqref{eq:defmu}, and \eqref{eq:defSigma} for $A$, $\mu$, and $\Sigma$ one finds
\begin{equation}
	\text{E}(\Delta)=\text{tr}(QW + QWQ\Sigma_1) + T^TWT,
\end{equation}
and
\begin{equation}
	\begin{aligned}
		\text{Var}(\Delta)&=2\text{tr}\left((QW + QWQ\Sigma_1)^2\right)\\
		&\qquad + 4T^TW(Q+Q\Sigma_1Q)WT.
	\end{aligned}
\end{equation}
Note that while $A$ and $\Sigma$ are matrices in data space, $Q$, $W$, $T$, and $\Sigma_1$ are all in parameter space and hence functions of the first two moments of $p_1$, $p_2$, and $q$ only. Finally, one finds for $\ere$ and $\sigma(D)$:
\begin{align}
	\ere &= \frac 1 2 \left( \text{tr}\left(\Sigma_1^{-1}\Sigma_2\right) - d - \log \frac {\det \Sigma_2} {\det\Sigma_1} +\text{E}(\Delta)\right),
	\label{eq:generalere}\\
	\sigma^2(D) &= \frac 1 4 \text{Var}(\Delta).
	\label{eq:sigmadgen}
\end{align}

To derive the quantities stated in section \ref{sec:the_gaussian_limit}, one can simply take two limits. When updating constraints with complementary data as in section \ref{sub:adding_supplementary_data2}, the relative entropy between prior and posterior is considered. In this case $q$ is identical with $p_1$, resulting in $T=0$ and $\Sigma_q^{-1} = \Sigma_1^{-1}$. Using these simplifications, one finds:
\begin{align}
	\ere &= -\frac 1 2 \log \frac {\det \Sigma_2} {\det\Sigma_1},\\
	\sigma^2(D) &= \frac 1 2 \text{tr}\left((\Sigma_p^{-1}\Sigma_{new} - \mathbb 1)^2\right).
\end{align}

When comparing the results of two datasets that replace each other as in section \ref{sub:substituting_data2}, the relative entropy between the two separately analyzed posteriors is of interest. Considering a wide prior for the derivation of $p_2$, all terms containing $\Sigma_q^{-1}$ are small compared to the terms independent of $q$, resulting in $Q\simeq\Sigma_2^{-1}$ and $T\rightarrow0$. One hence finds in this case:
\begin{align}
	\ere &\simeq -\frac 1 2 \log \frac {\det \Sigma_2} {\det\Sigma_1} + \text{tr}(\Sigma_2\Sigma_1^{-1}),\\
	\sigma^2(D) &\simeq \frac 1 2 \text{tr}\left((\Sigma_{old}^{-1}\Sigma_{new} + \mathbb 1)^2\right).
\end{align}

Using the results from section 3.1a in \cite{mathai1992quadratic}, equation \eqref{eq:genchi2} can be rewritten as a weighted sum of noncentral $\chi^2$ variables:
\begin{equation}
	\Delta = \sum_{i=1}^p\lambda_i Z_i,
\end{equation}
where $\lambda_i$ are the non-zero eigenvalues of $A\Sigma$ with multiplicity $n_i$ and $Z_i$ are noncentral $\chi^2$ variables with $n_i$ degrees of freedom and noncentrality parameter $\delta_i$. The noncentrality parameters $\delta=(\delta_1,\cdots,\delta_p)$ are given by $\delta = P\Sigma^{-\frac 1 2}\mu$ where $P$ is the orthogonal $p\times p$ matrix which diagonalizes $A\Sigma$. It can be shown that the non-zero eigenvalues of $A\Sigma$ are equal to the eigenvalues of $QW + QWQ\Sigma_1$, as both matrices are equivalent up to cyclic permutations. As $Q$, $W$, and $\Sigma_1$ are matrices in parameter space, the $\lambda_i$'s can be directly estimated from the moments of $p_1$, $p_2$, and $q$. The noncentrality parameters $\delta$ are zero when replacing data or analyzing complementary data, but non-zero for the most general case of partial replacement. They can however not be calculated from the moments of the parameter distributions and are therefore hard to estimate. As $\mu$ is small for the applications in section \ref{sub:numerical_results}, the influence of $\delta$ on the distribution of $D$ was simply neglected.

The p-value is the probability of $\Delta$ being greater or equal than the observed value. To estimate the p-values of the observed shifts, an algorithm by \citet{Davies:1980ty} was used, implemented in the R package \verb|CompQuadForm| by \citet{Duchesne:2010kq}. It needs the eigenvalues $\lambda_i$ and the observed shift $\Delta$ as an input and outputs an estimate for the p-value.



\section{Marginalizing Template Amplitudes in Gaussian Likelihoods} 
\label{sec:marginalizing_template_amplitudes}

\noindent Consider a likelihood that is normally distributed in some $d$-dimensional data D:
\begin{equation}
	\mathcal L(\Theta,A;D) = \mathcal N(D;M(\Theta,A),\Sigma),
\end{equation}
where the model is of the form $M(\Theta,A)=C(\Theta)+TA$ with $C(\Theta)$ being an arbitrary function taking $\Theta$ to a $d$-dimensional vector and $A$ being an $a$-vector of template amplitudes for the templates in $T$, a $d\times a$ matrix. Furthermore, assume that the prior on the amplitudes $A$ is normally distributed:
\begin{equation}
	p(A)=\mathcal N(A;A_A,\Sigma_A).
\end{equation}
The full posterior of $\Theta$ and $A$ is then given by Bayes rule $\p(\Theta,A)\propto \mathcal L(\Theta, A;D)p(A)p(\Theta)$. When the interest is in the marginalized posterior of $\Theta$ only, this implies that one wants to use the marginalized likelihood $\mathcal L(\Theta)\equiv\int dA\,p(A)\mathcal L(A)$ instead of the full likelihood for Bayes rule. This integral can be done in a similar spirit as in \ref{sub:deriving_the_posterior}, resulting in
\begin{equation}
	\mathcal L(\Theta;D) = \mathcal N(D;C(\Theta)+TA_A,\Sigma+T\Sigma_AT^T).
\end{equation}
The correcting factor on the covariance matrix and the model predictions can be easily implemented in the likelihood code for the data.

\section{Relative Entropy Estimation with Monte Carlo Methods} 
\label{sec:relative_entropy_estimation_with_monte_carlo_methods}

\noindent The numerical estimation of the relative entropy between posterior distributions is covered in the following.

\subsection{Gaussian approximation} 
\label{sub:gaussian_approximation}

\noindent In appendix \ref{sub:relative_entropy_between_gaussians} it is shown that the relative entropy between two normal distributions $p_1(X) = \mathcal N(X,\mu_1,\Sigma_1)$ and $p_2(X) = \mathcal N(X,\mu_2,\Sigma_2)$ is given by
\begin{equation}
	\begin{aligned}
		D(p_1||p_2) &= \frac 1 2 \left(\text{tr}(\Sigma_2^{-1}\Sigma_1) -d - \log \det (\Sigma_2^{-1}\Sigma_1) \right)\\
		&\qquad + \frac 1 2 (\mu_1 - \mu_2)^T \Sigma_2^{-1} (\mu_1 - \mu_2).
	\end{aligned}
	\label{eq:nnrelent}
\end{equation}
It is straightforward to estimate mean $\mu$ and covariance matrix $\Sigma$ of a distribution $p(X)$ from an MCMC sample $\{X_i\} = \{(x^{(1)}_i,\cdots,x^{(d)}_i)\}$:
\begin{align}
	\mu \equiv \langle X \rangle_p &\approx \frac 1 N \sum_{i = 1}^N X_i \equiv \bar X,\\
	\Sigma_{kl} \equiv \langle x^{(k)}x^{(l)} \rangle_p -\mu^{(k)}\mu^{(l)} &\approx \frac 1 N \sum_{i = 1}^N x^{(k)}_i x^{(l)}_i - \bar X^{(k)}\bar X^{(l)}.
\end{align}
So it is left to plug the resulting estimates for mean and covariance of prior and posterior into equation \eqref{eq:nnrelent} to find an estimate for the relative entropy gain. When one or both of the distributions depend on nuisance parameters, marginalization of these additional parameters is achieved by simply considering means and covariances of the cosmological parameters only. The MCMC samples for the values in Table \ref{tab:relent} were created with the \verb|CosmoHammer| package \cite{Akeret:2012uk}.


\subsection{General approach} 
\label{sub:general_approach}

\noindent In general, the relative entropy between two distributions $p_1$ and $p_2$ is given by the following integral:
\begin{equation}
	D(p_1||p_2) = \int d\Theta\, p_1(\Theta) \log \frac {p_1(\Theta)} {p_2(\Theta)}.
\end{equation}
When the distributions of interest are posteriors $p^{new}_1$ and $p^{new}_2$ from cosmological applications they can be numerically evaluated up to a normalization factor by calculating the product of prior $p_i$ and likelihood $\mathcal L_i$:
\begin{equation}
	\tilde p^{new}_i(\Theta) = \mathcal L_i(\Theta) p_i(\Theta) \propto p^{new}_i(\Theta).
\end{equation}
There are standard Monte Carlo techniques to evaluate expectation values, among them Monte Carlo Markov chains (MCMC), Monte Carlo integration, and nested sampling \cite{Skilling:ci}. Using one of those techniques, it is hence left to estimate normalization and relative entropy via
\begin{equation}
	N_i = \int d\Theta\, \tilde p_i^{new} = \langle \mathcal L_i \rangle_{p_i}
\end{equation}
and
\begin{equation}
	D(p^{new}_1||p^{new}_2) = \langle \log \frac {\tilde p^{new}_1} {\tilde p^{new}_2} \rangle_{p^{new}_1} + \log\frac {N_2} {N_1},
\end{equation}
where $\langle \;\cdot\; \rangle$ denotes the expectation values which have to be estimated. In this work, the \verb|CosmoHammer| package \cite{Akeret:2012uk} was used to estimate $\langle \log \frac {\tilde p^{new}_1} {\tilde p^{new}_2} \rangle$ from MCMC samples of $p^{new}_1$. Furthermore, a Monte Carlo integrator was employed to evaluate $\int d\Theta\, \tilde p^{new}_i$ over a five-sigma region of $p_i$, where the integral boundaries were estimated from covariance matrix and mean of the MCMC samples created for the estimation of $\langle \log \frac {\tilde p^{new}_1} {\tilde p^{new}_2} \rangle$.



\bibliography{paper}

\end{document}